# Norms and Commitment for iOrgs™ Information Systems: Direct Logic™ and Participatory Grounding Checking

Carl Hewitt
http://carlhewitt.info

iOrgs™ Information Systems raise important issues for formalizing norms that require extensions and revisions of previous foundational work. For example, extension and revision is required of the fundamental assumption of the Event Calculus:

> *Time-varying properties hold at particular time-points if they have been initiated by an action at some earlier time-point, and not terminated by another action in the meantime.*

The fundamental assumption of the Event Calculus is overly simplistic when it comes to organizations in which time-varying properties have to be *actively maintained and managed* in order to continue to hold and termination by another action is not required for a property to no longer hold. *I.e.,* if active measures are not taken then things will go haywire by default.

Similarly extension and revision is required for Grounding Checking properties of systems based on a set of ground inferences. Previously Model Checking has been performed using the model of nondeterministic automata based on states determined by time-points. These nondeterministic automata are not suitable for iOrgs, which are highly structured and operate asynchronously with only loosely bounded nondeterminism.

iOrgs Information Systems have been developed as a technology in which organizations have people that are tightly integrated with information technology that enables them to function organizationally. iOrgs formalize existing practices to provide a framework for addressing issues of authority, accountability, scalability, and robustness using methods that are analogous to human organizations. In general

- iOrgs are a natural extension Web Services, which are the standard for distributed computing and software application interoperability in large-scale Organizational Computing.
- iOrgs are structured by Organizational Commitment that is a special case of Physical Commitment that is defined to be information pledged.

iOrgs norms are used to illustrate the following:

- Even a very simple microtheory for normative reasoning can engender inconsistency In practice, it is impossible to verify the consistency of a theory for a practical domain.
- Improved Safety in Reasoning. It is not safe to use classical logic and probability theory in practical reasoning.

# Contents

# iOrgs™ Information Systems

***In the organization lies the power.***

iOrgs Information Systems have been developed as a technology in which organizations have people that are tightly integrated with information technology that enables them to function organizationally [Hewitt and Inman 1991; Hewitt 2008b, 2008d, 2008g].[i] iOrgs formalize existing practices to provide a framework for addressing issues of authority, accountability, scalability, and robustness using methods that are analogous to human organizations.[ii]. In general

- iOrgs are a natural extension Web Services, which are the standard for distributed computing and software application interoperability in large-scale Organizational Computing.
- iOrgs are structured by *Organizational Commitment* that is a special case of *Physical Commitment* [Hewitt 2007 2008b] that is defined to be *information pledged*.[iii]

This paper discusses how iOrgs require Direct Logic in reasoning and participatory grounding checking in systems analysis:

1. The development of iOrgs and the extreme dependence of our society on these systems have introduced new phenomena. These systems have pervasive inconsistencies among and within the following:
    - *Norms* that express how systems can be used and tested in practice
    - *Policies* that express over-arching justification for systems and their technologies
    - *Practices* that express implementations of systems

    Different parties are responsible for constructing, evolving, justifying and maintaining practices, norms, and operations for large-scale Organizational Computing. In specific cases any one consideration can trump the others. Sometimes debates over inconsistencies can become quite heated, *e.g.,* between sales, engineering and finance.
2. Grounding checking[iv] is a fundamental tool in the analysis of iOrgs. However, previous work on model checking has been performed using the model of nondeterministic automata based on states determined by time-points. These nondeterministic automata are not suitable for iOrgs, which are highly structured and operate asynchronously with only loosely bounded nondeterminism. Instead analysis based on regions of space-time (as in Participatory Semantics [Hewitt and Manning 1996]) is required.

## Participatory Semantics

Participatory Semantics [Hewitt and Manning 1996] is based on regions of space-time called Participations[v] according to the following legend:[vi]

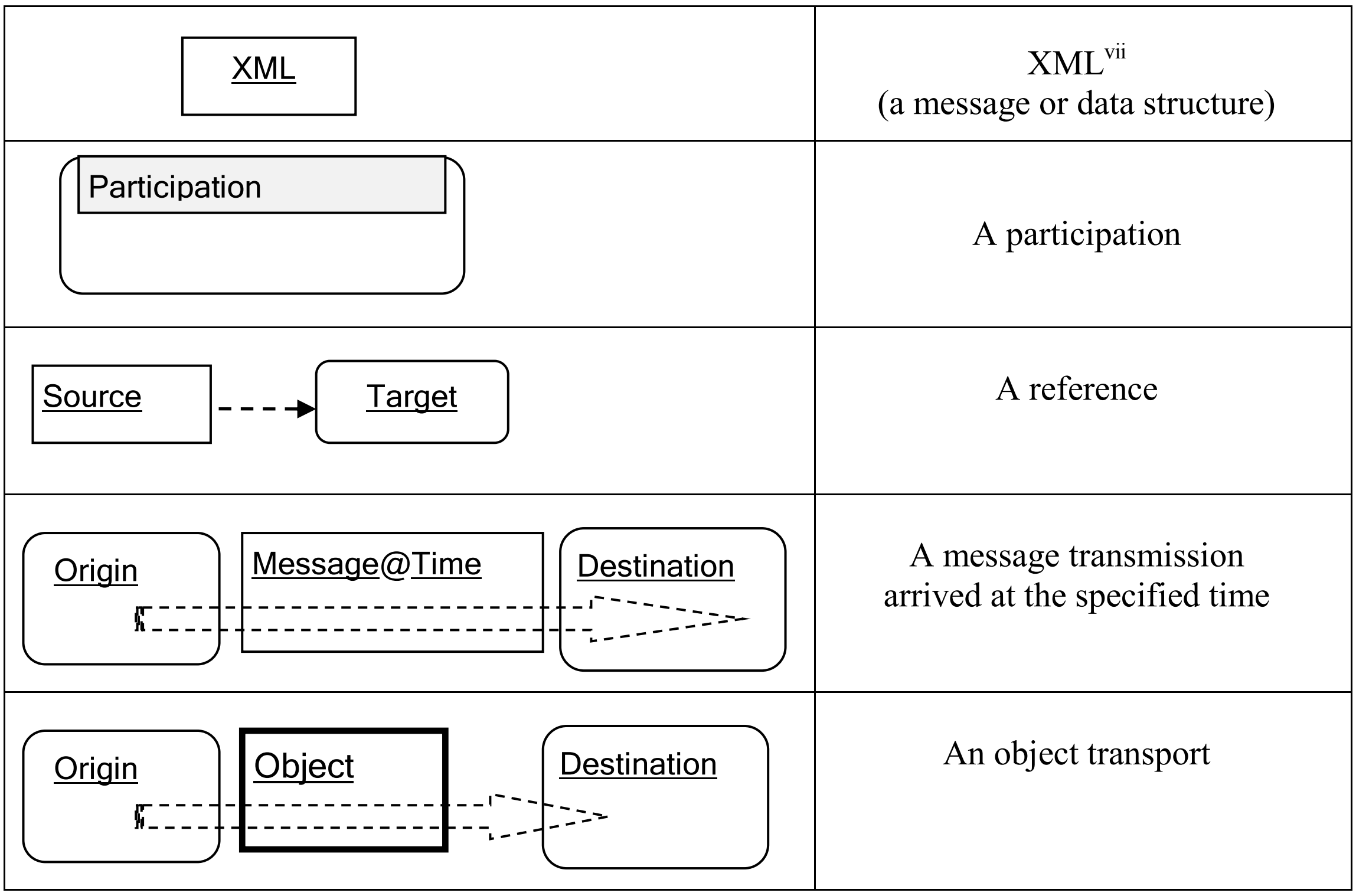

**Figure 1. Legend**

## Contrast between Participatory Semantics and the Event Calculus

Participatory Semantics is based on 4-dimensional regions of space-time whereas the Event Calculus is based on events (which take 0 time) on a global universal time-line.

This paper does not use the usual Event Calculus formalism [Kowalski and Sergot 1986, Miller and Shanahan 1999]. A principle reason for not adopting the Event Calculus is avoidance of its fundamental assumption:

> "Time-varying properties hold at particular time-points if they have been *initiated* by an action at some earlier time-point, and not *terminated* by another action in the meantime."

The fundamental assumption of the Event Calculus is overly simplistic when it comes to organizations in which time-varying properties have to be *actively managed* in order to continue to hold and termination by another action is not required for a property to no longer hold. *I.e.,* if active measures are not taken then things will go haywire by default. For example consider the following property: "*Drive safely"* It might be said that the property was "terminated" when a driver collides with another vehicle. However, it is often the case that some "unsafe driving" occurred before the collision!

Another problem with the Event Calculus is that it is formulated at the very low level of abstraction of time-points. As Carlo Rovelli has explained:[viii]

> *"We never really see time. We see only clocks. If you say this object moves, what you really mean is that this object is here when the hand of your clock is here, and so on. We say we measure time with clocks, but we see only the hands of the clocks, not time itself. And the hands of a clock are a physical variable like any other. So in a sense we cheat because what we really observe are physical variables as a function of other physical variables, but we represent that as if everything is evolving in time."*

Preoccupation with global time-points is a serious problem with the Event Calculus. This problem is closely related to another problem with the Event Calculus: A time is not bound to a locale but is instead imagined to be free floating!

Consider an example in which safe driving is followed by unsafe driving leading to a collision. An important issue is that there may be no event which clearly delineates the transition from safe driving to unsafe driving.[ix] The lack of such an event is not material to Participatory Semantics. However, there is no clear terminating event for the Event Calculus.

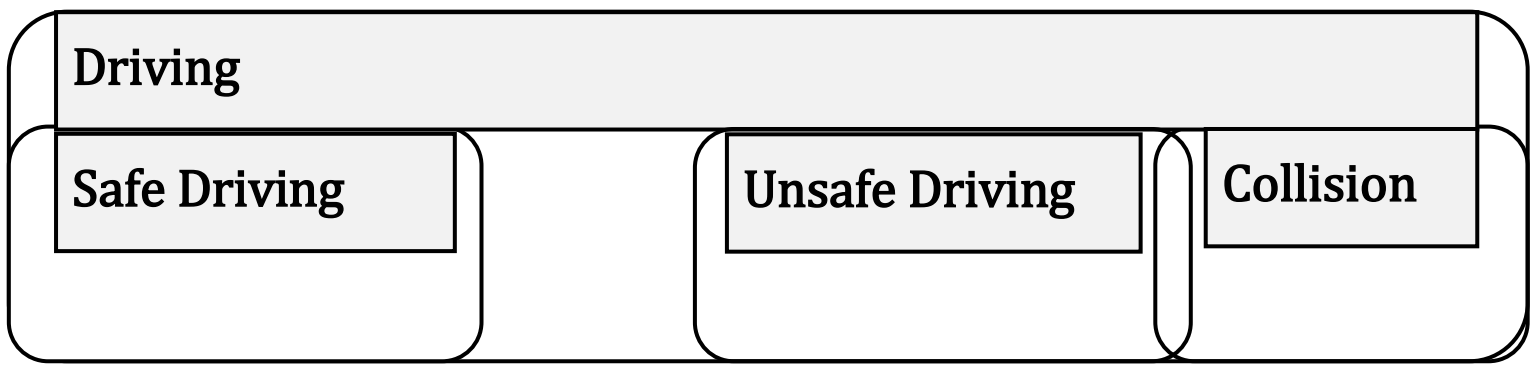

**Figure 2. Transition from Safe to Unsafe Driving (with no clipping event)**

Next consider an example in which AM transitions to PM on July 31, 2008 in California. Here the issue is that there is no *physical* event that occurs throughout California that marks the transition from AM to PM.

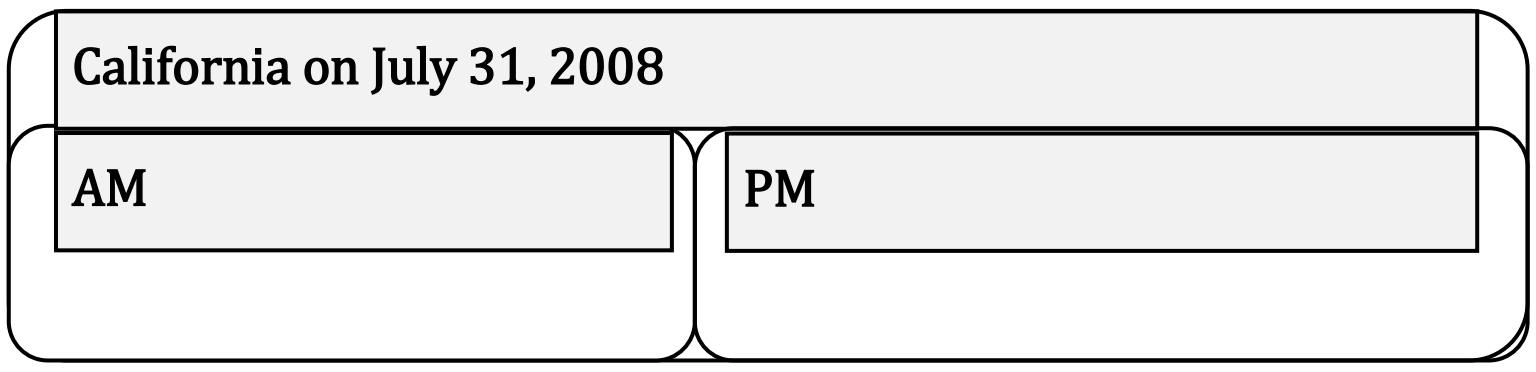

**Figure 3. Transition from AM to PM (with no clipping event)**

By convention, the AM and PM regions for California are adjacent to each other. However, this adjacency does not require the existence of any event that occurs throughout California and the lack of such an event is not material to Participatory Semantics. However, there is no terminating event as required by the Event Calculus.

## Commitment

According to [Hewitt 2007], a *Physical Commitment* PC is defined to be a *pledge* that certain *information* I holds for a *physical system* PS for a *space-time region* R. Note that physical commitment is defined for *whole physical systems*; not just a participant or process.

This paper uses a mythical *Santa Cruz FishMarket*[x] to illustrate how organizational commitments can be formulated at a higher level of abstraction. The Santa Cruz FishMarket uses an electronic English Auction starting with a reserved price in which a certain time is allowed for more bids to come in before the bidding is closed. As each higher bid is received, the new minimum bid is announced to the participants. Tie bids are broken by choosing the one which arrived first as the winner. Consequently the Santa Cruz FishMarket is an organizational commitment with an auction of buyers and sellers.

An implementation[xi] for a **SimpleAuction** for the Santa Cruz FishMarket is given in the appendix.[xii] The commitments below are made by the implementation pledging the following information:

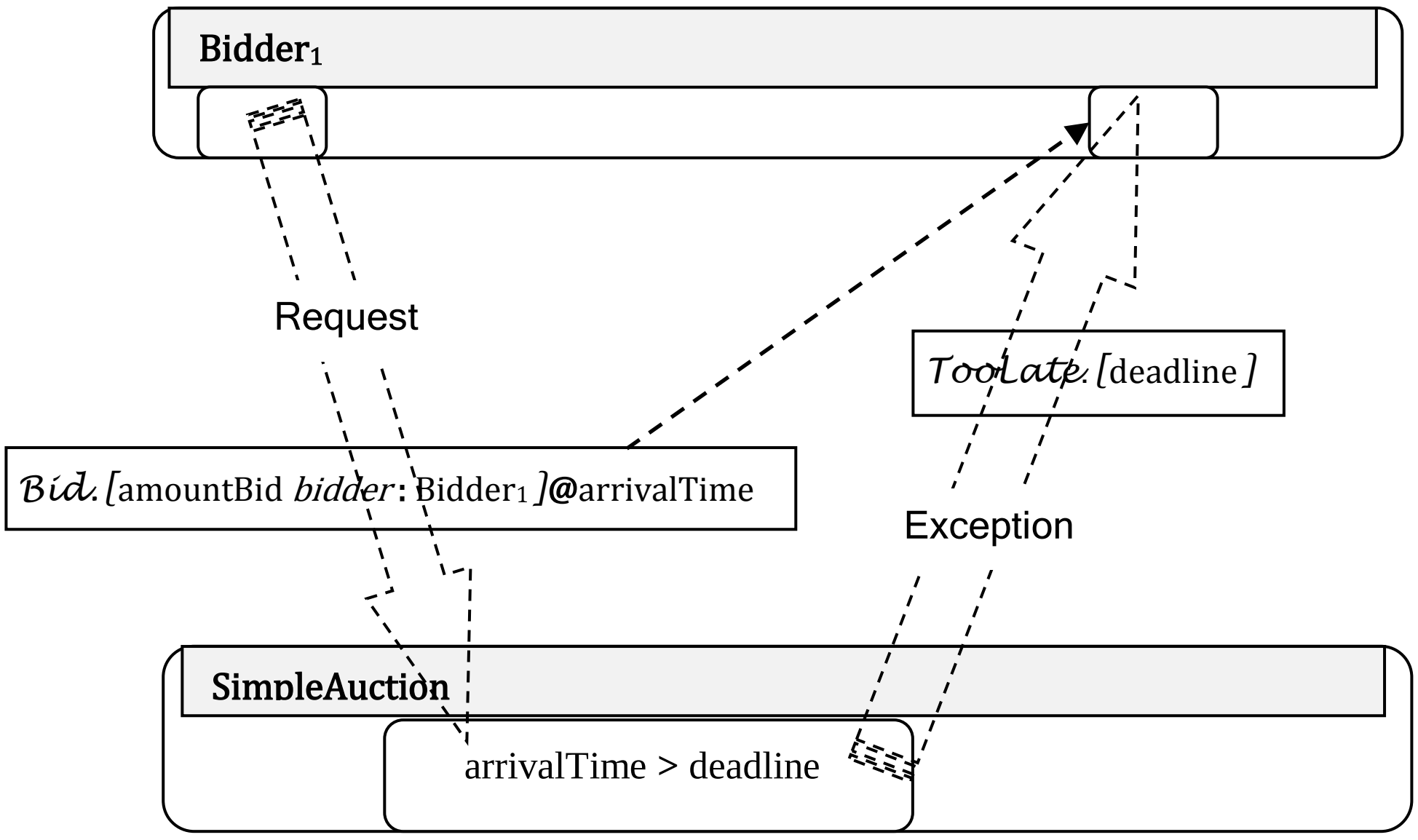

**Figure 4. Commitment: Bidding too late causes an exception**

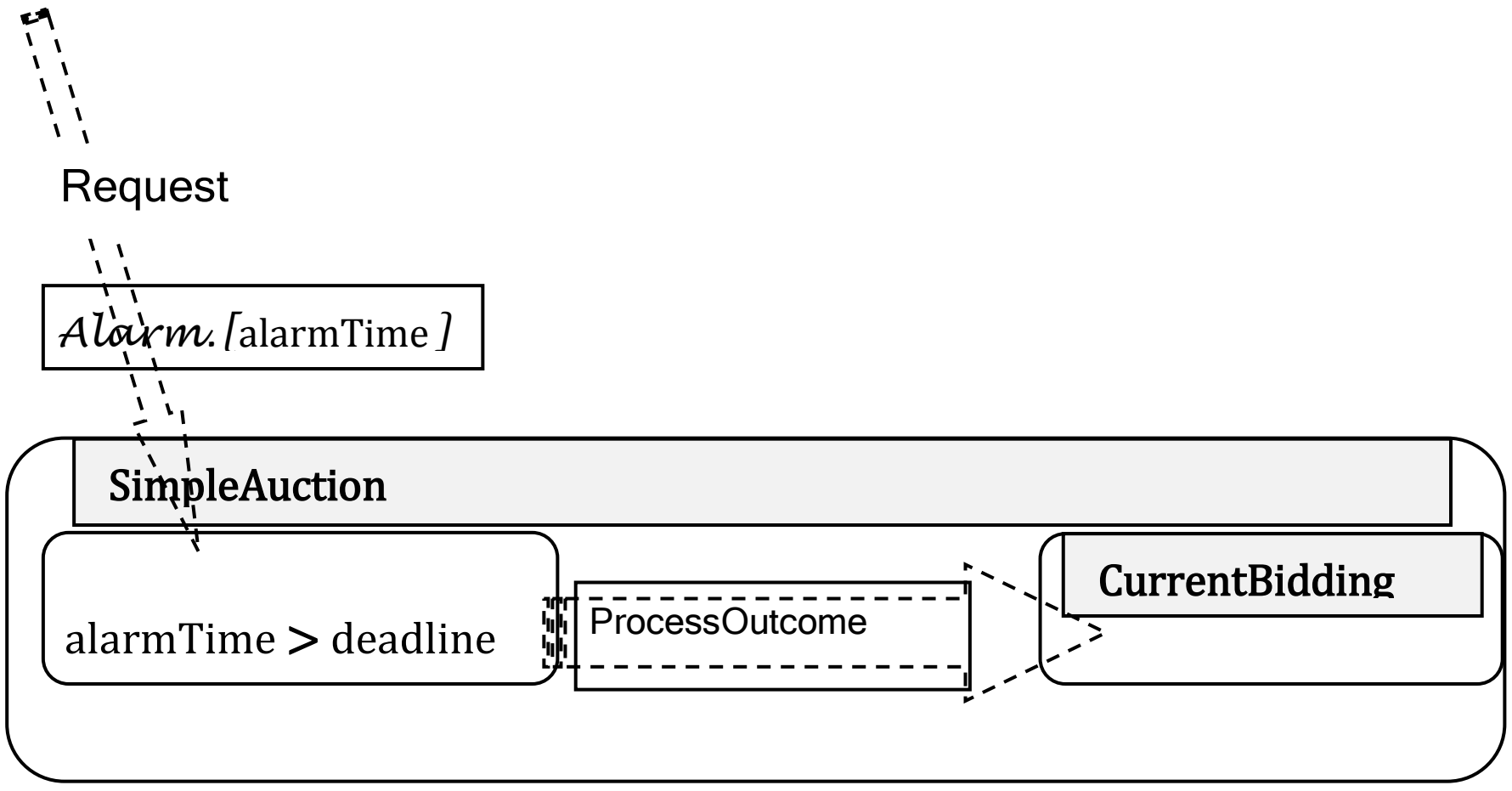

**Figure 5. Commitment: An alarm triggers the processing of the auction outcome**

## Participatory Grounding Checking

The denotational semantics of concurrency were first developed in [Clinger 1981]. Subsequently [Hewitt 2006b] developed the TimedDiagrams model with the ***Computational Representation Theorem*** which states:

The denotation $\mathrm{Denote}_S$ of a closed[xiii] system **S** represents all the possible behaviors of **S** as

$$\mathrm{Denote}_S = \bigsqcup_{i\in\omega} \mathrm{Progression}_S{}^{i}(\bot_S)$$

*where* $\mathrm{Progression}_S$ is an approximation function that takes a set of approximate behaviors to their next stage and $\bot_S$ is the initial behavior of **S**.

The denotational semantics exhibits relatively unbounded nondeterminism because in the delivery of a message can occur a relatively unbounded amount of time after it is sent. This relatively unbounded nondeterminism can cause trouble with traditional global state-based approaches [Bianculli, Morzenti, Pradella, San Pietro, Spoletini 2007; Bordini, Fisher, Visser, and Wooldridge 2004; Cliffe, De Vos, and Padget 2006; Desai, Cheng, Chopra, and Singh 2007; Venkatraman and Singh 1999, *etc*.] because of the following issues:

- State explosion because of the increase in possible interleavings of global states
- Modeling failure because the system being modeled is not finite state

Participatory grounding checking makes use of the Representation Theorem to characterize possible alternative computations. In participatory grounding checking:

- Explosion is less of a problem because local groundings are modeled instead of global state. Also Participatory Semantics can be used to abstract high level properties of denotations in a way that is similar to how abstraction has been used in global state model checking.

- Systems are not modeled as nondeterministic state machines, Petri nets, or process calculi [Aceto and Gordon 2005].[xiv]
- Communication is modeled as being fundamentally one-way and asynchronous. In this way modeling problems such as occur using Petri Nets and synchronous process calculi are avoided [Padget and Bradford 1998].

For example consider the system with **SimpleAuction** (defined in the appendix of this article) augmented with bidders like the following:[xv]

```
SimpleBidder ≡
 behavior {               ⓘ[xvi] serialize the messages received by this bidder
   Auction::theAuction          ⓘ auction that bidder is bidding for

   Currency::maximumBid         ⓘ maximum that bidder will bid for this auction
-----------------------------------------------------------------------------------------------
 implements Bidder              ⓘ the Bidder interface is implemented below
 self.newMinimum(amount) →
                                ⓘ a newMinimum message with amount has been received
   amount ??{
     (< maximumBid) :                          ⓘif amount is less than maximumBid
        theAuction.bid(amount for: self)            ⓘ then bid the minimum
          catcher {                                 ⓘ and if it throws the exception
            TooLittle[minimumBid] :                 ⓘ that the bid was too small
                relay self.newMinimum(minimumBid)}
                                                    ⓘ send yourself the new minimum bid
     not (< maximumBid) : void}}                    ⓘelse do nothing and return void
```

Using bidders like the above, execution scenarios of the system can be computed using the Representation Theorem. These execution scenarios can be checked against norms such as the following:

*Commitment:* ***At Santa Cruz FishMarket, deliveries from seller to buyer are paid at the agreed price; i.e. Santa Cruz FishMarket pledges the following information: For every*** **seller, buyer,** ***and*** **delivery,**

Delivers[seller, buyer, delivery] $\vdash_{\text{FishMarket}}$ PaysAgreedPrice[buyer, seller, delivery][xvii]

Note that seller, buyer, and delivery are all space-time participations in the above norm. Consequently, there is enough information to specify that the buyer pays the agreed price to the seller on delivery of the purchase.[xviii]

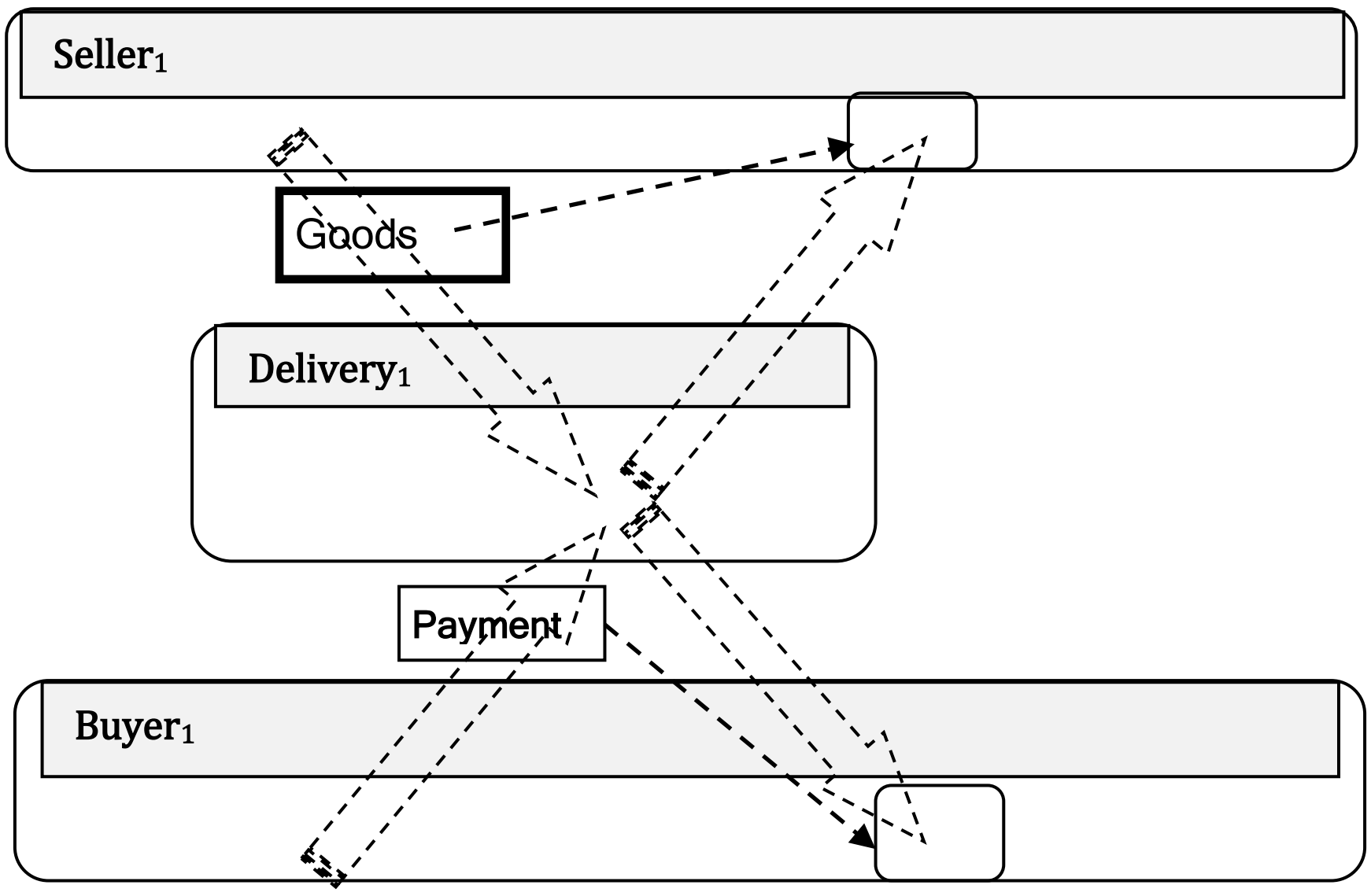

Figure 6. A commitment involving the exchange of Goods for Payment

## Direct Inference

Direct inference is used in to directly infer conclusions from premises. For example, suppose that we have

1) WeekdayAt5PM $\vdash_{\mathcal{Boston}}$ TrafficJam
which says that in *Boston*, a weekday at 5PM infers a traffic jam.

2) $\vdash_{\mathcal{Boston}}$ ¬TrafficJam
which says that in *Boston*, no traffic jam.

In classical logic, ¬WeekdayAt5PM is inferred in *Boston* from 1) and 2) above. But fortunately in Direct Logic:

$\nvdash_{\mathcal{Boston}}$ ¬WeekdayAt5PM

which says in *Boston*, there is a *particular* proposition (¬WeekdayAt5PM) that cannot be inferred in Direct Logic from 1) and 2) above.

Consequently, ***direct inference comes into play even in the absence of overt inconsistency.***[xix]

## Collusion at Santa Cruz FishMarket

Ethical conduct at the FishMarket is worthy of further study. A classic guide to norms for ethical conduct appears in *The Prince* [Machiavelli 1532]. So roughly in this spirit, consider the follow axiomatization[xx] of collusion at FishMarket:[xxi]

**1)** p, action: CanResult[Do[p, action], Rich[p]] $\vdash_{\mathit{ThePrince}}$ Do[p, action] ⓘ *axiom*[xxii]

**2)** $\vdash_{\mathit{ThePrince}}$ CanResult[Do[Machiavelli, Collude], Rich[Machiavelli]] ⓘ *axiom*

**2')** Do[Machiavelli, Collude] $\vdash_{\mathit{ThePrince}}$ CanResult[Rich[Machiavelli]]] ⓘ *axiom*

**3)** $\vdash_{\mathit{ThePrince}}$ **Do[Machiavelli, Collude]** ⓘ *from* 1) *and* 2)

**4)** p, action: CanResult[Do[p, action], Ruined[p]] $\vdash_{\mathit{ThePrince}}$ ¬Do[p, action] ⓘ *axiom*[xxiii]

**5)** $\vdash_{\mathit{ThePrince}}$ CanResult[Do[Machiavelli, Collude], Ruined[Machiavelli]] ⓘ *axiom*

**5')** Do[Machiavelli, Collude] $\vdash_{\mathit{ThePrince}}$ CanResult[Ruined[Machiavelli]]] ⓘ *axiom*

**6)** $\vdash_{\mathit{ThePrince}}$ **¬Do[Machiavelli, Collude]** ⓘ *from* 4), *and* 5)

Note that, in the *ThePrince*, there is an inconsistency between **3)** and **6)**. However, even though the *ThePrince* is inconsistent, it is not meaningless because in some respects it captures some underlying paradoxical aspects of Machiavelli's theory*:*

> *"The wish to acquire more is admittedly a very natural and common thing; and when men succeed in this they are always praised rather than condemned. But when they lack the ability to do so and yet want to acquire more at all costs, they deserve condemnation for their mistakes."*

## Inconsistent policy in War

A classic case of inconsistency occurs in the novel Catch-22 [Heller 1961] which states that a person "*would be crazy to fly more missions and sane if he didn't, but if he was sane he had to fly them. If he flew them he was crazy and didn't have to; but if he didn't want to he was sane and had to. Yossarian was moved very deeply by the absolute simplicity of this clause of Catch-22 and let out a respectful whistle. 'That's some catch, that Catch-22,' he observed.*"

In the spirit of Catch-22, consider the follow axiomatization of the above:

1. $\vdash_{\mathit{Catch\text{-}22}}$ ℙ(Able[x, Fly] ∧ ¬Fly[x]) ≤ ℙ(Sane[x]) ⓘ *axiom*
2. $\vdash_{\mathit{Catch\text{-}22}}$ ℙ(Sane[x]) ≤ ℙ(Obligated[x, Fly]) ⓘ *axiom*
3. $\vdash_{\mathit{Catch\text{-}22}}$ ℙ(Sane[x] ∧ Obligated[x, Fly]) ≤ ℙ(Fly[x]) ⓘ *axiom*
4. $\vdash_{\mathit{Catch\text{-}22}}$ ℙ(Fly[x]) ≤ ℙ(Crazy[x]) ⓘ *axiom*
5. $\vdash_{\mathit{Catch\text{-}22}}$ ℙ(Crazy[x]) ≤ ℙ(¬Obligated[x, Fly]) ⓘ *axiom*
6. $\vdash_{\mathit{Catch\text{-}22}}$ ℙ(Sane[p] ∧ ¬Obligated[p, Fly]) ≤ ℙ(¬Fly[p]) ⓘ *axiom*

For Yossarian, we have the following axioms:

**7.** $\vdash_{Catch\text{-}22} \mathbb{P}(\text{Able[Yossarian, Fly]}) \cong 1$ ⓘ *axiom*

**8.** $\vdash_{Catch\text{-}22} \mathbb{P}(\text{Sane[Yossarian]}) \cong 1$ ⓘ *axiom*

Consequently,

**2'.** $\vdash_{Catch\text{-}22} 1 \cong \mathbb{P}(\text{Obligated[Yossarian, Fly]})$ ⓘ Yossarian *using* **2** *and* **8**

**3'.** $\vdash_{Catch\text{-}22} 1 \cong \mathbb{P}(\text{Fly[Yossarian]})$ ⓘ Yossarian *using* **3**, **2'** *and* **8**

**4'.** $\vdash_{Catch\text{-}22} 1 \cong \mathbb{P}(\text{Crazy[Yossarian]})$ ⓘ Yossarian *using* **4** *and* **3'**

**5'.** $\vdash_{Catch\text{-}22} 1 \lesssim 1 - \mathbb{P}(\text{Obligated[Yossarian, Fly]})$ ⓘ Yossarian *using* **5** *and* **4'**

**5''.** $\vdash_{Catch\text{-}22} \mathbb{P}(\text{Obligated[Yossarian, Fly]}) \cong 0$ ⓘ *reformulation of* **5'**

**6'.** $\vdash_{Catch\text{-}22} 1 \lesssim 1 - \mathbb{P}(\text{Fly[Yossarian]})$ ⓘ Yossarian *using* **6'**, **8** *and* **5''**

**6''.** $\vdash_{Catch\text{-}22} \mathbb{P}(\text{Fly[Yossarian]}) \cong 0$ ⓘ *reformulation of* **6'**

Thus there is an inconsistency in *Catch-22* in that both of the following hold:

**3'.** $\vdash_{Catch\text{-}22} 1 \cong \mathbb{P}(\text{Fly[Yossarian]})$

**6''.** $\vdash_{Catch\text{-}22} \mathbb{P}(\text{Fly[Yossarian]}) \cong 0$

## Paradigm shift from Inconsistency Denial to Rapid Recovery

*ThePrince* and *Catch-22* illustrate the following important points:

- ***Even a very simple microtheory for normative reasoning can engender inconsistency*** In practice, it is impossible to verify the consistency of a theory for a practical domain.
- ***Improved Safety in Reasoning***. It is not safe to use classical logic and probability theory in practical reasoning.

## Norms at Santa Cruz FishMarket

Norms are commitment supported by communities of practice.

*Norm:* ***At Santa Cruz FishMarket, there is no collusion among buyers and sellers.***

Formalizing the norm above is the subject of future research.

## Conclusion

iOrgs raise important issues for inconsistency robustness and participatory grounding checking. This paper presents some ideas for formalizing these issues. Relationships among these issues are analyzed using illustrations from *FishMarket* and *The Prince*. The following conclusions are proposed.

- Extension and revision is required of the fundamental assumption of the Event Calculus: *Time-varying properties hold at particular time-points if they have been initiated by an action at some earlier time-point, and not terminated by another action in the meantime.* The fundamental assumption of the Event Calculus is overly simplistic when it comes to iOrgs in which time-varying properties have to be *actively maintained and managed* in order to continue to hold and termination by another action is not required for a property to no longer hold. *I.e.,* if active measures are not taken then things will go haywire by default. Consequently the Event Calculus approach must evolve into a strongly paraconsistent system structured around participations in space-time.
- Similarly extension and revision is required for Model Checking properties of systems. Previously Model Checking has been performed using the model of nondeterministic automata based on states determined by time-points. These nondeterministic automata are not suitable for iOrgs, which are highly structured and operate asynchronously with only loosely bounded nondeterminism. Consequently Model Checking needs to evolve in the direction of verifying participatory behavior in iOrgs.

## Acknowledgments

The development of Participatory Semantics was joint work with Carl Manning. Les Gasser, Mike Huhns, Victor Lessor, Pablo Noriega, Sascha Ossowski, Jaime Sichman, Munindar Singh, *etc.* provided valuable suggestions at AAMAS'07. The reviewers and participants of MALLOW'07 (including John Lloyd, John-Jules Meyer, Pablo Noriega, Jaime Sichman, Munindar Singh, Rineke Verbrugge, *etc.*) provided valuable comments. Afterwards Munindar Singh provided helpful pointers to the literature. The reviewers for COIN@AAMAS'08 made helpful suggestions. Conversations with Jeremy Forth were helpful in developing the comparison of Participatory Semantics and the Event Calculus. Munindar Singh made extensive comments and suggestions that significantly improved the presentation. A helpful review was provided by an anonymous referee for the special issue of the Journal of IGPL on Normative Multi-agent Systems.

## Bibliography

- Chris Anderson. *The End of Theory: The Data Deluge Makes the Scientific Method Obsolete* Wired. June 23, 2009.
- Luca Aceto and Andrew D. Gordon (Ed.) *Algebraic Process Calculi: The First Twenty Five Years and Beyond* 2005
- Marco Alberti, Marco Gavanelli, Evelina Lamma, Paola Mello, and Paolo Torroni "Modeling interactions using social integrity constraints: A resource sharing case study" *DALT*. LNAI 2990. Springer-Verlag, 2004.
- Aldo Antonelli. "Non-monotonic Logic" *Stanford Encyclopedia of Philosophy*. March 2006.

- Domenico Bianculli, Angelo Morzenti, Matteo Pradella, Pierluigi San Pietro, Paola Spoletini "Trio2Promela: A Model Checker for Temporal Metric Specifications" *ICSE'07*.
- Brian Blum. *Contracts: Examples and Explanations* Aspen Publishers. 3rd edition 2004.
- Rafael Bordini, Michael Fisher, Willem Visser, and Michael Wooldridge "Verifiable Multi-Agent Programs" Springer LNAI 3067. 2004.
- Geof Bowker, Susan L. Star, W. Turner, and Les Gasser. (Eds.) *Social Science Research, Technical Systems and Cooperative Work*. Lawrence Earlbaum. 1997.
- Cristiano Castelfranchi "Commitments: From individual intentions to groups and organizations" *AAAI-93 Workshop on AI and Theories of Groups and Organizations: Conceptual and Empirical Research*.
- Amit Chopra and Munindar Singh. *Multiagent Commitment Alignment*. Draft of March 7, 2009.
- Owen Cliffe, Marina De Vos, and Julian Padget (2006). "Specifying and Reasoning about Multiple Institutions" *COIN@AAMAS'06.*
- Jiangbo Dang, Jingshan Huang and Mike Huhns "Workflow Coordination for Service-Oriented Multi-agent Systems" *AAMAS'07*.
- Nirmit Desai, Zhengang Cheng, Amit Chopra, and Munindar Singh "Toward Verification of Commitment Protocols and their Compositions" *AAMAS'07*.
- Virginia Dignum *A Model for Organizational Interaction*. PhD thesis. Utrecht. 2004..
- Rogier Van Eijk, Frank De Boer, Wiebe Van Der Hoek and John-Jules Meyer. "A verification framework for agent communication" *AAMAS'03*.
- Andrew Farrell, Marek Sergot, Mathias Salle and Claudio Bartolini "Using the event calculus for tracking the normative state of contracts" *International Journal of Cooperative Information System* June-September 2005.
- Tim Folger. "Newsflash: Time May Not Exist" Discover. June 12, 2007.
- Dorian Gaertner, Pablo Noriega, J.-A. Rodriguez-Aguilar, Wamberto Vasconcelos (2007). "Distributed Norm Management in Regulated Multi-Agent Systems" *AAMAS'07.*
- Joseph Heller. *Catch-22* Simon and Schuster. 1961.
- Carl Hewitt and Jeff Inman (1991). "DAI Betwixt and Between: From 'Intelligent Agents' to Open Systems Science" *IEEE Transactions on Systems, Man, and Cybernetics*. Nov. /Dec. 1991.
- Carl Hewitt and Carl Manning (1996). *Synthetic Infrastructures for Multi-Agency Systems ICMAS'96*.
- Carl Hewitt *What is Commitment? Physical, Organizational, and Social (Revised)* Pablo Noriega *.et. al.* editors. LNAI 4386. Springer-Verlag. 2007.
- Carl Hewitt (2008a) *Large-scale Organizational Computing requires Unstratified Reflection and Strong Paraconsistency* Coordination, Organizations, Institutions, and Norms in Agent Systems III. Jaime Sichman, Pablo Noriega, Julian Padget and Sascha Ossowski (Ed.). Springer-Verlag. 2008. *http://organizational.carlhewitt.info/*
- Carl Hewitt (2008b). *ORGs for Scalable, Robust, Privacy-Friendly Client Cloud Computing* IEEE Internet Computing September/October 2008.
- Carl Hewitt (2008c) *Middle History of Logic Programming: Resolution, Planner, Edinburgh Logic for Computable Functions, Prolog and the Japanese Fifth Generation Project* *ArXiv 0904.3036*
- Carl Hewitt (2008d) *A historical perspective on developing foundations for client cloud computing: iConsult™ & iEntertain™ Apps using iInfo™ Information Integration for iOrgs™ Information Systems* *(Revised version of "Development of Logic Programming: What went wrong, What was done about it, and What it might mean for the future" AAAI Workshop on What Went Wrong. AAAI-08.) ArXiv 0901.4934.*
- Carl Hewitt (2008e) *Common sense for concurrency and inconsistency tolerance using Direct Logic™ and the Actor Model* *ArXiv 0812.4852.*

- Carl Hewitt (2008f) *Scalable Privacy-Friendly Client Cloud Computing: a gathering Perfect Disruption* Stanford Computer Systems Laboratory Colloquium Oct. 22, 2008. Video recording. *http://stanford-online.stanford.edu/courses/ee380/081022-ee380-300.asx*
- Carl Hewitt (2008g) *Perfect Disruption: Causing the Paradigm Shift from Mental Agents to ORGs* IEEE Internet Computing. Jan/Feb 2009.
- Carl Hewitt *ActorScript™: Industrial strength integration of local and nonlocal concurrency for Client-cloud Computing* ArXiv. 0907.3330
- Bryan Horling and Victor Lesser (2005). "Using ODML to Model Multi-Agent Organizations" IAT'05
- Nicholas Jennings "Commitments and conventions: The foundation of coordination in multi-agent systems" *Knowledge Engineering Review*. 3. 1993.
- Frederick Knabe "A Distributed Protocol for Channel-Based Communication with Choice" *PARLE'92*
- Bill Kornfeld and Carl Hewitt "The Scientific Community Metaphor" *IEEE Transactions on Systems, Man, and Cybernetics*. January 1981.
- Robert Kowalski and Marek Sergot. "A Logic-based Calculus of Events" *New Generation Computing* Volume 4 , Issue 1. 1986.
- Robert Kowalski. "Database updates in the Event Calculus" *Journal of Logic Programming*. 1992.
- Niccolò Machiavelli (1532) *The Prince* (Bantam Classics 1984).
- Robin Milner "Elements of interaction: Turing award lecture." *CACM*. January 1993.
- Pablo Noriega *Agent Mediated Auctions: The Fishmarket Metaphor*. Ph.D. Barcelona. 1997.
- Julian Padget and Russell Bradford (1998). "A π-calculus Model of a Spanish Fish Market" *AMET-98* also in *LNAI 1571* Springer Verlag. 1999.
- Carl Petri *Kommunikation mit. Automate*. Ph. D. Thesis. University of Bonn. 1962.
- Murray Shanahan and Rob Miller (1999) "The event-calculus in classical logic — alternative axiomatization," *Electronic Transactions on Artificial Intelligence* 3(1).
- Jaime Sichman, Virginia Dignum, and Cristiano Castelfranchi (2005). "Agent organizations." *JBCS*, 11(3).
- Munindar Singh. "Social and Psychological Commitments in Multi-agent Systems" *AAAI Fall Symposium on Knowledge and Action at Social and Organizational Levels*. 1991
- Munindar Singh "An ontology for commitments in multi-agent systems: Toward a unification of normative concepts" *Artificial Intelligence and Law* 7. 1999.
- Feng Wan and Munindar Singh (2005). "Formalizing and achieving multiparty agreements via commitments" *AAMAS'05*.
- Michael Winikoff, Wei Liu, and James Harland (2005). "Enhancing commitment machines" *DALT*. LNAI 3476. Springer-Verlag. 2005.
- Pinar Yolum and Munindar Singh "Enacting Protocols by Commitment Concession" *AAMAS'07*.
- Mahadevan Venkatraman and Munindar Singh. "Verifying Compliance with Commitment Protocols: Enabling Open Web-Based Multi-agent Systems" JAAMAS. 1999.

## Appendix 1. A simple auction procedure in ActorScript

```
SimpleAuction ≡
 behavior {
   Bidders::theBidders              ⓘ a collection of those allowed to bid on this auction
   Currency::minimumBid               ⓘ current minimum bid for this auction
   Time::deadline                 ⓘ current deadline by which this auction will end unless
                                ⓘ another higher bid is received for this auction
   Bidding::currentBidding            ⓘ a recording of the current state of bidding for this auction
---------------------------------------------------------------------------------------------
 implements Auction             ⓘ the Auction interface is implemented below
 self.bid(amountBid)@Time::arrivalTime →
        ⓘ a message with amount bid and bidder has been received at arrivalTime
   arrivalTime ??{      ⓘ if arrivalTime
     (> deadline) : throw  TooLate.[deadline];  ⓘ is after deadline, complain bid is too late
     not(> deadline) :
       amountBid ?? {                          ⓘelse if the amount bid is smaller than the minimum
         (< minimumBid) : throw TooLittle.[minimumBid];  ⓘ then complain the bid is too little
         not(< minimumBid) :
               {currentBidding.bid(amountBid time: arrivalTime);  ⓘ record the bid in currentBidding
                         ⓘ this may throw an exception if the bidder is unqualified
               let(Time::newDeadline = CurrentTime()+δ,
                   Amount::newMinimumBid = amountBid * 110%)
                              ⓘ compute the new deadline and new minimum bid
                    {theBidders.newMinimum(newMinimumBid deadline: newDeadline),
                       ⓘ inform the allowed bidders of the new minimum and deadline

                     self.alarm(newDeadline),  ⓘ set an alarm for this auction with a new deadline
                     Acknowledgment.[ ] also become (minimumBid = newMinimumBid,
                                                  deadline = newDeadline)}}}}
                      ⓘ return  acknowledgment that the bid has been accepted and
                    ⓘ also update this auction with the new minimum bid and deadline
 alarm.(alarmTime) →  ⓘ an alarm message with alarmTime has been received
   alarmTime ?? {                   ⓘif alarmTime
     (<deadline) : void;                ⓘ is before the deadline return void
     not(<deadline) : currentBidding.processOutcome}}
    ⓘ else return the result of  processing the outcome of this auction according to the currentBidding
```

## End Notes

[i] The architecture of an iOrg differs fundamentally from a Mental Agent that cognitively operates in a human-like fashion. The Mental Agent paradigm [Alberti, Gavanelli, Lamma, Mello, and Torroni 2004] has had some success in modeling and simulating human-like behavior. However, computing has changed dramatically from the time of its invention and we are in the midst of a "perfect disruption" [Hewitt 2008g] brought on by the following:

- *Hardware*. Many-core architecture that will soon support thousands of threads in a process for widely-used software applications using semantic integration (see below).
- *Software*. Client cloud computing in which information is permanently stored in servers on the Internet and cached temporarily on clients that range from single chip sensors, handhelds, notebooks, desktops, and entertainment centers to huge data centers. (Even data centers are clients that often cache their information to guard against geographical disaster.) Client cloud computing will provide much needed new capabilities including the following:
  - maintaining the privacy of client information by storing it on servers encrypted so that it can be decrypted only by using the client's private key. (The information is unencrypted only when cached on clients.)
  - providing greater integration of user information obtained from servers of competing vendors without requiring them to interact with each other.
  - providing better advertising relevance and targeting without exposing client privacy.
- *Applications*. Scalable semantic integration, e.g., integrating the following:
  - calendars and to do lists
  - email archives
  - presence information including physical, psychological and social
  - documents (including presentations, spread sheets, proposals, job applications, photos, videos, gift lists, memos, purchasing, contracts, articles, *etc*.)
  - contacts (including social graphs)
  - search results
  - marketing and advertising relevance influenced by the above

This perfect disruption is causing a paradigm shift from Mental Agents to iOrgs as the foundation for implementing large-scale Internet applications [Hewitt 2008g].

[ii] For background information on iOrgs see [Bowker, Star, Turner, and Gasser 1977; Dignum 2004, Singh and Huhns 2005; Horling and Lesser 2005].

[iii] In some previous work, the subject of contracts [Blum 2004, *etc*.] has been treated using the (somewhat unfortunate name) "commitment" for contractual obligations [Singh 1991, Jennings 1993; Noriega 1997; Singh and Huhns 2005; Chopra and Singh 2009]. In this paper, these obligations are treated as special cases of Physical Commitment. See Hewitt [2007].

[iv] *i.e*., checking the behavior of a system against a model

[v] Note that Participations (being regions of space-time) represent *both* objects and activities.

[vi] Note that Participatory Semantics is based on space-time as opposed to the more usual approach of basing semantics just on time, *e.g.* the Event Calculus [Farrell, Sergot, Salle, and Bartolini 2005], *etc.*

[vii] XML is used be because it is increasingly dominant as the *de facto* standard for structured message communication and stands to become the *de facto* standard for data structures.

[viii] quoted in [Folger 2007]

[ix] For example, such an event would have to exist in the Event Calculus formulation in Kowalski [1992].

[x] Inspired by the Blanes FishMarket Metaphor [Noriega 1997].

[xi] appendix in the concurrent programming language ActorScript [Hewitt 2008c]

[xii] A functional notation is used for XML. For example
**PersonName<First<"Kurt"> Last< "Gödel">>** can print as:
<PersonName> <First> Kurt </First> <Last> Gödel </Last> </PersonName>

[xiii] Closed means that the system does not receive any communications from outside itself.

[xiv] The Actor Model subsumes other models of concurrency, *e.g.* Process Calculi [Milner 1993; Aceto and Gordon 2005] and Petri Nets [Petri 1962] using a two-phase commit protocol [Knabe 1992].

[xv] Implemented in ActorScript™ [Hewitt 2010]

[xvi] The symbol ⓘ is used to begin a comment that extends to the end of the line.

[xvii] Expressed in Direct Logic [Hewitt 2008c] (see discussion later in this paper).

[xviii] Note that (unlike Venkatraman and Singh [1999]), no assumption is made that the buyers and sellers are not malicious, *e.g.*, no use is made of time-stamps that can be forged.

[xix] Statistical probabilistic (fuzzy logic) systems are affected follows: Suppose (as above)

$$\vdash_{\mathcal{Boston}} \mathbb{P}(\text{TrafficJam} \mid \text{WeekdayAt5PM}) = 1$$ [xix]

$$\vdash_{\mathcal{Boston}} \mathbb{P}(\text{TrafficJam}) = 0$$

Then

$$\vdash_{\mathcal{Boston}} \mathbb{P}(\text{WeekdayAt5PM}) = \frac{\mathbb{P}(\text{WeekdayAt5PM} \wedge \text{TrafficJam})}{\mathbb{P}(\text{TrafficJam} \mid \text{WeekdayAt5PM})} = 0$$ [xix]

Thus contraposition is built into probabilistic (fuzzy logic) systems and consequently incorrect information can be generated.

The above example illustrates that the choice of how to incorporate measurements into statistics can effectively determine the model being used. In this particular case, the way that measurements were taken did not happen to take into account things like holidays and severe winter storms This point was largely missed in [Anderson 2008] which stated

*"Correlation is enough."* ***We can stop looking for models. We can analyze the data without hypotheses about what it might show.*** *We can throw the numbers into the biggest computing clusters the world has ever seen and let statistical algorithms find patterns where science cannot."* (emphasis added)

[xx] The axiomatization makes use of higher order capabilities. For example a predicate like CanResult can take arguments Do[Machiavelli, Collude] and Rich[Machiavelli] to form the proposition CanResult[Do[Machiavelli, Collude], Rich[Machiavelli]].

[xxi] The axiomatization uses a colon (**:**) to separate universally quantified variable from the following proposition to which they apply.

Direct Logic supports fine grained reasoning because inference does not necessarily carry argument in the contrapositive direction**.** For example, the general principle "A person does anything that can make them rich"

(*i.e.,* CanResult[Do[p, action], Rich[p]] $\vdash_{\mathit{ThePrince}}$ Do[p, action] does not support the inference of ¬CanResult[Do[Machiavelli, Collude], Rich[Machiavelli]] from ¬Do[Machiavelli, Collude]). *E.g.,* it might be the case that CanResult[Do[Machiavelli, Collude], Rich[Machiavelli]] even though it infers Do[Machiavelli, Collude] contradicting ¬Do[Machiavelli, Collude].

[xxii] "*A prince never lacks legitimate reasons to break his promise"* [Machiavelli 1532]

[xxiii] "*The one who adapts his policy to the times prospers, and likewise that the one whose policy clashes with the demands of the times does not."* [Machiavelli 1532]